\documentclass[12pt]{article}
\usepackage{amsmath,epsfig,amssymb,eufrak}
\usepackage{amscd,latexsym}
\usepackage{color}

\date{}

\def\xlf{\raisebox{+0.2em}{\color{red}\boldmath{$\chi$}}\hspace{-0.2ex}\raisebox{-0.2em}{\color{green}L}
\hspace{-1.5ex}\raisebox{+0.14em}{\color{blue}F}\hspace{2mm}}

\def\lsi{\raise0.3ex\hbox{$<$\kern-0.75em\raise-1.1ex\hbox{$\sim$}}}
\def\gsi{\raise0.3ex\hbox{$>$\kern-0.75em\raise-1.1ex\hbox{$\sim$}}}

\newcommand{\gsim}{\mathop{\gsi}}

\input macros.sty      % \eqalign, \meqalign
\input eqalign.sty     % \eqalign, \meqalign

\begin{document}

\begin{titlepage}
\title{
  {\vspace*{-1cm} \normalsize
  \hfill \parbox{40mm}{DESY/04-207\\
                       HU-EP-04/63
                       SFB/CPP-04-57}}\\[10mm]
Going chiral: overlap versus twisted mass fermions}  
\author{W.~Bietenholz$^{\, 1}$, 
S.~Capitani$^{\, 2}$,
T.~Chiarappa$^{\, 3}$, 
N.~Christian$^{\, 3,4}$, \\
M.~Hasenbusch$^{\, 5}$, 
K.~Jansen$^{\, 3}$, 
K.-I.~Nagai$^{\, 3}$,
M.~Papinutto$^{\, 3}$, \\
L.~Scorzato$^{\, 1}$,
S.~Shcheredin$^{\, 1}$, 
A.~Shindler$^{\, 3}$,
C.~Urbach$^{\, 3,4}$,\\
U.~Wenger$^{\, 3}$ and
I.~Wetzorke$^{\, 3}$\\ 
\\
   {\bf \xlf Collaboration}\\
\\
{\small $^{1}$ Institut f\"{u}r Physik, Humboldt Universit\"{a}t zu Berlin} \\
{\small Newtonstr.\ 15, D-12489 Berlin, Germany}\\\vspace*{-0.4cm}\\
{\small $^{2}$ Institut f\"{u}r Physik/Theoretische  Physik}\\ 
{\small Universit\"{a}t Graz, A-8010 Graz, Austria} \\ \vspace*{-0.4cm}\\
{\small $^{3}$   NIC/DESY Zeuthen} \\ 
{\small         Platanenallee 6, D-15738 Zeuthen, Germany}
\\\vspace*{-0.4cm}\\
{\small $^{4}$  Institut f\"{u}r Theoretische Physik, Freie Universit\"{a}t Berlin} \\
{\small Arnimallee 14, D-14195 Berlin, Germany}\\\vspace*{-0.4cm}\\
{\small $^{5}$ Dipartimento di Fisica, Universit\`a di Pisa}\\
{\small Via Buonarroti 2, I-56127 Pisa, Italy}
}

\maketitle

\begin{abstract}
We compare the behavior of overlap fermions, which are chirally 
invariant, and of Wilson twisted mass fermions at full twist in the approach 
to the chiral limit. Our quenched simulations reveal that 
with both formulations of lattice fermions pion masses of 
$\mathcal{O}$(250 MeV) can be reached in practical applications.
Our comparison is done at a fixed value of the lattice spacing $a\simeq
0.123$ fm. A number of quantities are measured such as hadron masses,
pseudoscalar decay constants and quark masses obtained from 
Ward identities. We also determine the
axial vector renormalization constants in the case of overlap fermions.

\vspace{0.8 cm}
\noindent
%%%{\it PACS:}  ; ; \\
%%%{\it Keywords:}  ; .
\end{abstract}

\end{titlepage}

\section{Introduction}

Approaching the so-called physical point in lattice QCD, where the light 
meson masses
assume their values observed in nature, is a major challenge for 
numerical simulations in
lattice field theory. 
In this regime of QCD, where the
light quarks, i.e.\ the $u$ and $d$ flavors, play the dominant role,
the explicit breaking
of chirality on the lattice, the role of topological charge excitations and
the appearance of unphysical very-low lying eigenvalues of the lattice Dirac 
operator can render simulations very demanding, if not impossible.

The complicated low energy dynamics of 
light quarks of the fundamental theory can be effectively described by 
{\em chiral perturbation theory} \cite{GaLe-p}. 
In chiral perturbation theory ($\chi$PT) an effective Lagrangian is constructed
with terms that are compatible with
the global symmetries of the original QCD Lagrangian
and that are ordered
according to some hierarchy, which depends on the expansion regime. In this
effective Lagrangian, each term has a coefficient which enters as a
free parameter 
that cannot be determined within the chiral
perturbation theory itself.
Such coefficients are denoted as the {\em low energy
constants},
and they play a vital role in many QCD processes at low energy.
The challenge is then to determine these low energy constants from first
principles, i.e.\ directly from QCD as the underlying,
fundamental theory.

Such a link between the effective 
theory and fundamental QCD   
can, in principle, be provided by
lattice techniques.
However, simulations at values of quark masses, where such a contact 
to $\chi$PT can safely be made, are very difficult, as 
mentioned above.
Standard approaches using the Wilson fermion discretization
of lattice QCD are confronted with technical algorithmic problems 
since 
in this case the bare quark mass does not provide an infrared regulator.
In this paper we will consider 
two formulations of lattice QCD
that are able to overcome this problem: overlap fermions \cite{HN}
and twisted 
mass fermions \cite{tm}. In particular, we will concentrate on 
the Wilson twisted mass (Wtm)    
\cite{Frezzotti:2003ni} formulation of lattice fermions
at a full twisting angle of $\pi /2$.
Both, overlap and Wtm fermions lead to 
$\mathcal{O}(a)$ improvement for physical quantities 
and are expected to 
allow for simulations at very small
quark masses, corresponding to their physical values
as estimated from experiment. Overlap fermions induce an exact lattice
chiral symmetry \cite{ML} and provide a sound definition 
of the topological charge \cite{HLN}. Hence they
have conceptual advantages, but 
they are, unfortunately, rather expensive in 
numerical simulations.
Twisted mass fermions on the other hand are rather cheap 
to simulate but show residual chiral symmetry breaking effects.
Hence it is not clear how the benefits of both formulations of
lattice QCD compare in practical simulations 
and here we would like to give a first 
direct comparison. 

We compute a number of quantities, Ward identity quark masses, 
meson and baryon masses, decay constants and renormalization factors, 
driving the values of quark masses small enough that
$\chi$PT is expected to be applicable. 
While for Wtm fermions this is the first work along this line, we refer
to
Refs.~\cite{others} for other simulations in the $p$-regime of chiral 
perturbation theory in
the case of overlap fermions.

We emphasize that the present comparison between both lattice fermions 
is only performed in the quenched approximation at 
$\beta=5.85$, corresponding to a lattice spacing of
$a\simeq 0.123$ fm. No attempt of a scaling analysis is performed here, see
however Ref.~\cite{JSUW} for a first work in this direction for Wtm. It is
the main goal of this paper to investigate how both formulations of 
lattice QCD behave in their approach to the chiral limit. In particular we
are aiming at an investigation 
to how small quark masses both formulations can be driven when used in
practical simulations.

We will also provide a computing time
estimate from our results in Ref.~\cite{solver}.
This question is most important for eventual dynamical simulations. If, 
for many
quantities, Wilson twisted mass fermions can reach quark masses 
that can be compared 
with overlap fermions, their advantage in the simulation cost will 
immediately benefit in dynamical simulations. For first results of 
dynamical twisted mass fermions, see Ref.~\cite{tmdyn}.

\section{Overlap and Wilson twisted mass fermions}

In this section we discuss the two fermion formulations of lattice
QCD that we have employed to study the approach to the chiral limit
at small values of the quark mass: overlap fermions and Wilson twisted
mass fermions. 
We emphasize some relevant points for the present work and refer to
Refs.~\cite{overlap,wtmqcd} for further discussions. 

\subsection{Overlap fermions}

Over the last few years a lattice fermion formulation 
leading to an exact lattice chiral symmetry  
was elaborated, namely the {\em Ginsparg-Wilson fermions}. 
The corresponding lattice Dirac operator $D_{\rm GW}$ satisfies the 
Ginsparg-Wilson relation \cite{GW}
\be  \label{GWR}
D_{\rm GW} \gamma_{5} + \gamma_{5} D_{\rm GW} = 2 a D_{\rm GW} \gamma_{5} R
D_{\rm GW} \ ,
\ee
where $R$ is a local term. 
The realization of an operator $D_{\rm GW}$ that we use here is the overlap
fermion, which is characterized by the Neuberger-Dirac operator \cite{HN}.
For $R_{xy}= \delta_{xy}/(2\rho)$ it takes the form
\bea
D_{\rm ov} &=& \Big( 1 - \frac{m_{\rm ov}\bar a}{2} \Big) D_{\rm ov}^{(0)} 
+ m_{\rm ov} \ , \nn \\
D_{\rm ov}^{(0)} &=& \frac{1}{\bar a} 
\left\{ 1 + A / \sqrt{ A^{\dagger} A} \right\} 
\ , \quad
A = a D_{\rm W} - \rho \ ,  \label{overlap}
\eea
where $\bar a\equiv a/\rho$ and $D_{\rm W}$ is the standard Wilson-Dirac operator, 
\be
D_{\rm W} = \sum_{\mu=1}^4 \frac{1}{2} 
[ \gamma_\mu(\nabla_\mu^* + \nabla_\mu) - a \nabla_\mu^*\nabla_\mu].
\label{Dw}
\ee
$\nabla_\mu$ and $\nabla_\mu^*$ denote the usual forward
and backward covariant lattice derivatives, $m_{\rm ov}$ is the bare quark mass
and $\rho \gsim 1$ is a mass parameter, which we set to 1.6.

$D_{\rm ov}^{(0)}$ (the overlap operator at zero quark mass)
obeys the Ginsparg-Wilson relation in Eq.~(\ref{GWR}).
It does obey a lattice modified but
exact chiral symmetry, which turns into the standard chiral symmetry
in the continuum limit \cite{ML}. This symmetry protects the lattice fermion
from additive mass renormalization and from $\mathcal{O}(a)$ lattice
artifacts, i.e. the action built by the operator in
Eq.~(\ref{overlap}) is $\mathcal{O}(a)$ improved.
It also implies that there are
exact zero modes with a definite chirality \cite{HLN}. Thus the topological 
charge can be identified as the index obtained from these zero modes.
$\mathcal{O}(a)$ improved bilinears are constructed as follows:
\be
O^{\rm ov}_\Gamma=\bar \psi^\alpha \Gamma \Big(1-\frac{\bar a
  D_{\rm ov}^{(0)}}{2}\Big)\psi^\beta=\frac{1}{1-\frac{\bar a m_\beta}{2}}\left(\bar
\psi^\alpha \Gamma \psi^\beta\right)
\ee
where $\psi^\alpha$ and  $\psi^\beta$ are two different flavours, and 
the last equality holds for correlation functions at non-zero physical 
distance. 
In the following we will use the following notation: $O^{\rm ov}_{\gamma_5}\equiv P$, 
$O^{\rm ov}_{\rm I}\equiv S$, $O^{\rm ov}_{\gamma_\mu}\equiv V_\mu$ and  
$O^{\rm ov}_{\gamma_\mu\gamma_5}\equiv A_\mu$.  

By now the overlap fermion has a very well established theoretical
basis, but its simulation is rather tedious. In our code the inverse
square root is approximated by Chebyshev polynomials to an absolute accuracy
of $10^{-15}$ (see Ref.~\cite{solver} for details). 
On this level of precision the lattice
chiral symmetry is certainly reliable. Unfortunately the computational
effort exceeds the one for Wtm by a large amount,  
which means that, at least for large volumes, for the time being 
only {\em quenched}
QCD simulations are possible with Ginsparg-Wilson fermions, at least when 
chiral symmetry is to be realized to the precision enforced in this paper. 
The virtues of this formulation 
include also a protection against exceptional configurations, thus allowing 
simulations at small pion masses, certainly when they are comparable or larger
than their physical values.

\subsection{Wilson twisted mass fermions}

As an alternative to regulate exceptionally small eigenvalues, 
Wilson fermions with a
{\em twisted mass} \cite{tm,Frezzotti:2003ni} can be used. 
This means that the Wilson-Dirac
operator obtains a mass term of the form
\be
m_{\rm tm} + i \mu \gamma_{5} \tau^{3} \ ,
\ee
where $m_{\rm tm}$ is again the bare quark mass, $\tau^{b}$ are the usual
Pauli matrices acting in flavor space and $\mu$ is the ``twisted mass''.

In this paper we will work with Wilson twisted mass fermions that can be 
arranged to be $\mathcal{O}(a)$ improved without  
additional improvement terms. To be more precise, 
 let us start by writing the Wtm QCD action (in the twisted basis) as
\begin{equation}
  \label{tmaction}
  S[U, \psi  , \bar\psi ] = a^4 \sum_x \bar\psi(x) ( D_W + m_{\rm tm} + i \mu
\gamma_5\tau^3 ) \psi(x)\; ,
\end{equation}
where the operator $D_W$ is given in Eq.~(\ref{Dw}).

The action as it stands in Eq.~(\ref{tmaction}) can, of course, be
studied in the full parameter space $(m_{\rm tm},\mu)$.
A special case arises, however, when $m_{\rm tm}$ is tuned towards a critical
bare quark mass $m_{\mathrm{c}}$.
In this, and only in this situation, all physical quantities are (or
can easily be) $\mathcal{O}(a)$ improved.
It is hence natural to rewrite
\be
\label{m0}
m_{\rm tm}=m_{\mathrm{c}} + \widetilde{m}
\ee
with $\tilde{m}$ an offset quark mass.
The values of $m_{\mathrm{c}}$ need only to be known with
$\mathcal{O}(a)$ accuracy \cite{Frezzotti:2003ni} and can be, for
instance, taken from the pure Wilson theory at $\mu=0$.

In this approach there is no need of improving the operators, so we will
consider the usual local bilinears 
\bea
 P^b=\bar \psi \gamma_5 \frac{\tau^b}{2}\psi &\quad&
 S^b=\bar \psi \frac{\tau^b}{2}\psi
 \nonumber\\
 A^b_\mu=\bar \psi \gamma_\mu\gamma_5 \frac{\tau^b}{2}\psi &\quad&
 V^b_\mu=\bar \psi \gamma_\mu\frac{\tau^b}{2}\psi
\eea
where $b$ is a $SU(2)$ flavour index.

Of particular interest is the PCVC relation.
In the twisted basis it takes the form
\be
\partial_\mu^* V^b_\mu  =  -2\mu\epsilon^{3bc}P^c ,  
\ee
where $\partial_\mu^*$ is the lattice backward derivative.
Through a vector variation of the action one obtains the point-split
vector current as defined in Ref.\ \cite{tm}.
This current is protected against renormalization and using the point-split 
vector current, the PCVC relation is an exact
lattice identity.
This implies that $Z_P=Z_\mu^{-1}$, where 
$Z_\mu$ is the renormalization constant for the twisted mass 
$\mu$. This will become important in the extraction of
the pseudoscalar decay constant $f_{\pi}$ as described below. 
Recent tests have 
revealed a very promising scaling behavior of this lattice fermion formulation
(in quenched simulations) \cite{JSUW}. 

\section{Numerical results}

Using standard heat bath and over-relaxation techniques to generate
gauge field configurations with the Wilson plaquette action, we have
performed various simulations at $\beta=5.85$, which corresponds to a value
of the lattice spacing $a\simeq 0.123$ fm ($a^{-1}\simeq 1.605$ GeV using 
$r_0 = 0.5$ fm \cite{r0}). Periodic boundary conditions were used in this 
work.

For overlap fermions we have 140 configurations on a $12^3\times 24$ 
lattice ($L_{12}\sim 1.48$ fm). The bare quark
masses are $m_\mathrm{ov} a=0.01,0.02,0.04,0.06,0.08,0.10$.
The simulations for twisted mass fermions are done at 
full twist. This is achieved by choosing $m_\mathrm{tm}=m_c$ in the 
pure Wilson theory at $\mu=0$. In the 
standard hopping parameter $\kappa$ notation
the offset quark mass is related to $\kappa$ by 
$\widetilde m = \frac{1}{2\kappa} - \frac{1}{2 \kappa_{\rm c}}$. 
Choosing $m_\mathrm{tm}=m_c$ then corresponds to use 
$\kappa = \kappa_{\rm c}=0.16166(2)$ from the vanishing of the 
pion mass for Wilson fermions 
\cite{JSUW}.
The twisted quark mass parameter was chosen to be 
$\mu a=0.005,0.01,0.02,0.04,0.06,0.08,0.10$. We then accumulated 
140 configurations on $12^3\times 24$, 
140 configurations on $14^3\times 32$ ($L_{14}\sim 1.72$ fm) and
380 configurations on $16^3\times 32$ ($L_{16}\sim 1.97$ fm)
lattices.
In tables and plots below both $m_\mathrm{ov}$ and $\mu$ are called
$m_\textrm{bare}$.
 
Our simulations, performed at a number of bare quark mass values,
applied a multiple mass solver (MMS) in both cases. The MMS for twisted mass
fermions will be described in a forthcoming paper~\cite{solver}, 
whereas the MMS for the usual fermion
formulations can be found in Ref.\ \cite{MMS}.
In order to improve the projection on the fundamental states of our
correlators, we have implemented the Jacobi smearing in the
way it is described in Refs.\ \cite{Baxter,Allton:1993wc}. 

In the following we will extract several physical quantities 
(hadron masses and decay constants) using the overlap and
the Wtm formulation of lattice QCD. 
Since the flavor structure of the quark propagators is important in the Wtm
case, we will specify the flavor content of the local operators only for
this formulation. We will indicate the fermionic action used in order
to distinguish the correlation functions, where necessary, of the two
formulations.

\subsection{Meson masses}

The first quantities we compared are the meson masses
$M_\textrm{meson}$. They are extracted by fitting the suitable two-point
correlation functions to the standard expression obtained from a spectral
decomposition and retaining only the fundamental state ($a \ll x_0 \ll T$)
\noindent 
\be
a^3 \sum_{\vec x} \langle O^\dag(\vec x, x_0) O(0)\rangle = 
\frac{|\langle 0|O|\textrm{meson}\rangle|^2 }
{M_\textrm{meson}}e^{-M_\textrm{meson}\frac{T}{2}}\cosh\left[
M_\textrm{meson}\left(x_0 - \frac{T}{2}\right)\right]
\ee
in the time interval $[t_\textrm{min},\frac{T}{2}]$ 
(with $a \ll t_\textrm{min} < \frac{T}{2}$). $t_\textrm{min}$ 
has been chosen by considering the effective mass, the
dependence of the fit on $t_\textrm{min}$ and by comparing with a
two-state fit. We extract the pseudoscalar masses (both degenerate
and non-degenerate) from the correlation functions 
\be
C_{P,\rm tm}^b(x_0) = a^3 \sum_{\vec x} 
\langle P^b(\vec x , x_0 )P^b(0)\rangle_{\rm tm} \ , \quad b=1,2\; .
\ee 
For overlap fermions we consider 
\be
C_{P,\rm ov}(x_0) = a^3 \sum_{\vec x} \langle P^\dag(\vec x, x_0 )
P(0) \rangle_{\rm ov}\; ,
\ee
and
\be
C_{P-S,\rm ov}(x_0) = a^3 \sum_{\vec x} \langle
P^\dag(\vec x , x_0 )P(0)-S^\dag(\vec x , x_0)S(0) \rangle_{\rm ov}\; ,
\ee
where the contribution of the topological zero modes 
cancels \cite{fvzeromode,others}. 
This last method has the drawback that the scalar meson appears as an
excited state and can affect the extraction of the mass of the ground state
for large quark masses. In our study, however, the quark masses
are sufficiently small such that this problem never occurs. 

\begin{table}[!t]
\begin{center}
\begin{tabular}{ccccccc}
\hline
\hline
$m_{\rm bare} a$ & $M^{P}_{\pi,\textrm{ov}}a$ &
$M^{P-S}_{\pi,\textrm{ov}}a$ & $M^{12^3\times24}_{\pi,\textrm{tm}}a$  & 
$M^{14^3\times32}_{\pi,\textrm{tm}}a $ & $M^{16^3\times32}_{\pi,\textrm{tm}}a $ 
&$M^{16^3\times32}_{\pi,\textrm{tm}} L_{16}$ \\
\hline
0.005 &    -     &     -      &     -       &     -       & 0.1700(25) & 2.7\\
0.01 & 0.212(9) & 0.140(20) & 0.2327(70) & 0.2301(37) & 0.2254(19) & 3.6\\
0.02 & 0.237(7) & 0.196(14) & 0.3193(48) & 0.3175(30) & 0.3122(16) & 5.0\\
0.04 & 0.299(5) & 0.280(10) & 0.4520(40) & 0.4506(23) & 0.4452(14) & 7.1\\
0.06 & 0.355(4) & 0.346(8)  & 0.5596(35) & 0.5575(19) & 0.5535(12) & 8.9\\
0.08 & 0.405(4) & 0.401(7)  & 0.6541(31) & 0.6510(17) & 0.6488(11) & 10.4\\
0.10 & 0.450(4) & 0.451(6)  & 0.7417(26) & 0.7378(16) & 0.7359(11) & 11.8\\
\hline
\hline
\end{tabular}
\end{center}
\caption{\it We show numerical results for the pion
masses obtained with overlap fermions using both the correlators $C_{P,\rm ov}$ 
and, $C_{P-S,\rm ov}$, and with Wtm fermions using three different volumes.
The overlap results were obtained on a $12^{3} \times 24$ lattice, and
we set $\beta = 5.85$ everywhere. The bare quark mass $m_{\rm bare}$ 
corresponds to $m_\mathrm{ov}$ in the overlap and to 
$\mu$ in the twisted mass case.}
\label{tab:mmesons1}
\end{table}

\begin{table}[!b]
\begin{center}
\begin{tabular}{cccccc}
\hline
\hline
$m_{\rm bare} a$& $M^{P-S}_{\pi,\textrm{ov}}a$ &
$M^{P-S}_{\pi,\textrm{ov}} L_{12}$ & $M^{12^3\times24}_{\pi,\textrm{tm}}a$  &
$M^{14^3\times32}_{\pi,\textrm{tm}}a$  &
$M^{16^3\times32}_{\pi,\textrm{tm}}a$ \\
%$M^{16^3\times32}_{\pi,\textrm{tm}} L_{16}$\\
\hline
0.01 & 0.134(22)& 1.6 & 0.2332(70) & 0.2303(39) & 0.2257(21)\\
0.02 & 0.192(16)& 2.3 & 0.3203(49) & 0.3172(28) & 0.3126(18) \\
0.04 & 0.275(12)& 3.4 & 0.4523(41) & 0.4493(22) & 0.4455(16) \\
0.06 & 0.342(10) & 4.2 & 0.5584(38) & 0.5564(20) & 0.5538(14)\\
0.08 & 0.397(10) & 4.8 & 0.6510(36) & 0.6504(19) & 0.6492(13) \\
0.10 & 0.448(8) & 5.4 & 0.7353(35) & 0.7363(18) & 0.7361(13) \\
\hline
\hline
\end{tabular}
\end{center}
\caption{\it The same as Table \ref{tab:mmesons1} but with sink smearing.}
\label{tab:mmesons2}
\end{table}

\begin{figure}
\vspace{-0.0cm}
\begin{center}
\epsfig{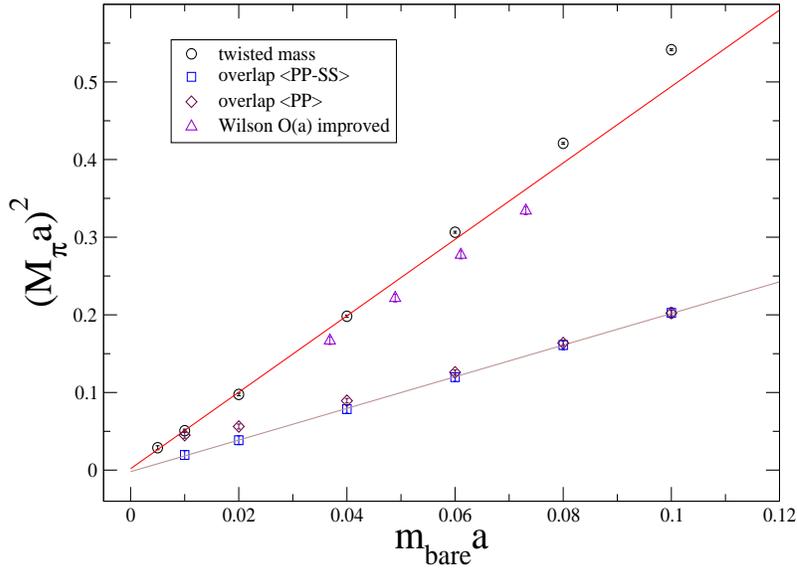}
\end{center}
\vspace{-0.0cm}
\caption{ \label{fig:mpi_ov_tm}
{\it Comparison of quenched results for the pion mass squared as a function
of the bare quark mass for three lattice fermions: standard $\mathcal{O}(a)$ improved
Wilson fermions \cite{HJLW01,HJLWproc}, twisted mass fermions and overlap fermions.
The bare quark mass corresponds to $m_\mathrm{ov}$ in the overlap and to 
$\mu$ in the twisted mass case.}}
\end{figure}

In Figure \ref{fig:mpi_ov_tm} and Tables \ref{tab:mmesons1},~\ref{tab:mmesons2} 
we give an overview of our results
for the pseudoscalar meson masses. For twisted mass fermions we have performed
simulations on three volumes, such that we can investigate finite volume 
effects. On the $12^3\times 24$ lattice, for the largest  bare quark
masses there are small contaminations from the excited states 
(which include also states of opposite parity appearing as 
$\mathcal{O}(a^2)$ artifacts) which are removed by using sink-smeared
correlators. 
Apart from this effect, small finite volume effects are visible
at the smallest masses. An analysis along  the lines of Ref.~\cite{fvme} 
shows that, on the pion masses corresponding to smallest value of 
$M_\pi L$ (i.e. those corresponding to $\mu a=0.01$ for $L=L_{12}$ 
and to $\mu a=0.005$ for $L=L_{16}$),  the finite volume effects
are within 2-3 percent and at most within two standard deviations 
from the extrapolated infinite volume limit. 
In practice they are thus not really relevant for the following discussion.
Below, we will present only those Wtm results which were
obtained on the $16^3\times32$ lattice, for which finite volume effects 
are completely negligible as long as $\mu a$ is larger than 0.005.

The problem of isolating the ground state does not appear
in the overlap case when we analyze the $C_{P,\rm ov}(x_0)$ correlator.
Even when we introduce a parity violating term explicitly with the 
$C_{P-S,\rm ov}(x_0)$ correlator, we still do not have any problem
to extract the pion mass for all the values of the quark masses simulated.
This is due to the fact that, at a fixed value of the bare quark mass, 
the pseudoscalar masses are smaller (and the gap between the 
fundamental and the excited states in the pseudoscalar correlator or between
the fundamental state of the pseudoscalar and of the scalar correlators 
larger) than the corresponding masses for Wtm.

Given the value of $M_{\pi}L$ in the case of overlap fermions and the
experience from Wtm fermions, we 
expect very small finite volume effects for the 5 heaviest quark 
masses (at the level of few percents in the case of $m_{\rm ov}a=0.02$). 
For the lowest quark mass (for which $M_{\pi,\mathrm{ov}} L=1.6$) finite
volume effects can be more relevant and thus we usually do not include the
corresponding data point in the fits. However, the analysis of the
various quantities presented below suggests that, for this quark mass, 
finite volume effects are not larger than our statistical error.

Note that for $\mathcal{O}(a)$ improved Wilson fermions
results \cite{HJLW01,HJLWproc}, which are also 
shown in Figure~\ref{fig:mpi_ov_tm}, the simulations had to be stopped at 
rather large values of the quark mass to avoid the appearance of
exceptional configurations. On the contrary, with both 
Wilson twisted mass fermions and overlap fermions we can reach very low
values of the quark and hence of the pion mass. 
 
For overlap fermions, the masses extracted from $C_{P-S,\rm ov}(x_0)$
have, to a very good approximation, a linear behavior down to the
smallest mass ($M_\pi\simeq220$ MeV). A linear extrapolation to the chiral 
limit gives an intercept of $-0.002(6)$. 
For Wilson twisted mass fermions we have performed two
fits: a linear one on the four smallest masses from which we get 
an intercept of $0.0017(2)$
with $\chi^2/d.o.f.=1.35$, and a quadratic one on all of the 7 points 
from which we
get an intercept of $0.0045(4)$ with $\chi^2/d.o.f.=0.19$. 
As we can see, the data show a
behavior which is much better described by a parabola than
by a straight line. We attribute the value of the intercept, 
non-compatible with zero in the chiral
limit, to the $\mathcal{O}(a)$ error in 
$\kappa_c$, which at $\mu=0$ gives an $\mathcal{O}(a)$
residual pion mass. The dependence of $M_\pi^2$ upon $\mu$ and
$a$ at small values of the twisted mass can in principle be computed in
$\chi$PT with the inclusion of the twisted mass term~\cite{Luigi}.  
The curvature of the Wtm data at high values of $\mu$, 
absent for overlap fermions,
can be explained by the uncertainty of $\mathcal{O}(a)$ in $\kappa_c$ 
(determined in the pure Wilson case $\mu=0$). This induces
an uncertainty of $\mathcal{O}(a^2\mu^2)$ in the pion mass, 
which increases
with a higher value of the twisted mass. 

\begin{figure}
\vspace{0.7cm}
\begin{center}
\epsfig{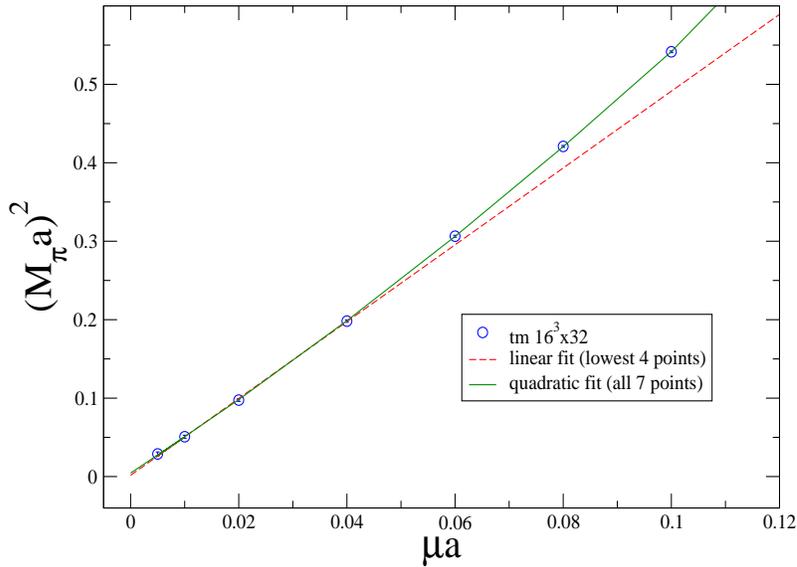}
\end{center}
\vspace{-0.0cm}
\caption{\it \label{fig:mpi_tm_lin_quad}
Twisted mass fermions: linear and quadratic fit of the squared pion masses
as function of the bare quark mass.}
\end{figure}

We clearly observe that the pion masses obtained with twisted mass fermions
shown in Figure \ref{fig:mpi_ov_tm}
always lay above the ones from the simulations using overlap fermions. 
This suggests that the renormalization factor $Z_{\rm m}$ of the quark mass 
should be larger for Wilson twisted mass fermions when compared to 
overlap fermions. It also means that smaller values of the quark mass 
have to be simulated with Wilson twisted mass fermions to reach the 
same pion mass as with overlap fermions. \\

As a next quantity we consider the {\em vector meson mass}. 
It has been extracted from the following correlators,
\begin{eqnarray*}
C_{A,\rm tm}^b(x_0) &=& \frac{a^3}{3}
\sum_{k=1}^3\sum_{\vec x} \langle A_k^b(\vec x, x_0 )A_k^b(0)\rangle_{\rm tm} \ , 
\quad b=1,2 \ , \\
C_{V,\rm ov}(x_0) &=& \frac{a^3}{3}
\sum_{k=1}^3\sum_{\vec x} \langle V_k^\dag(\vec x, x_0 )V_k(0)\rangle_{\rm ov} \ . 
\end{eqnarray*}

\begin{table}
\begin{center}
\begin{tabular}{ccc}
\hline
\hline
$m_{\rm bare} a$& $M^{\textrm{ov}}_{\rho} a$ &
$M^{\textrm{tm}}_{\rho} a$ \\
\hline
0.005 &   -     & 0.356(48) \\
0.01 & 0.632(34)& 0.468(46) \\
0.02 & 0.638(26)& 0.543(24) \\
0.04 & 0.653(16)& 0.6482(91) \\
0.06 & 0.666(12)& 0.7333(57)\\
0.08 & 0.683(9) & 0.8087(41)\\
0.10 & 0.702(8) & 0.8799(33)\\
\hline
\hline
\end{tabular}
\end{center}
\caption{\it Vector meson masses with sink smearing.}
\label{tab:mmesons3}
\end{table}

\begin{figure}[!t]
\vspace{-0.0cm}
\begin{center}
\epsfig{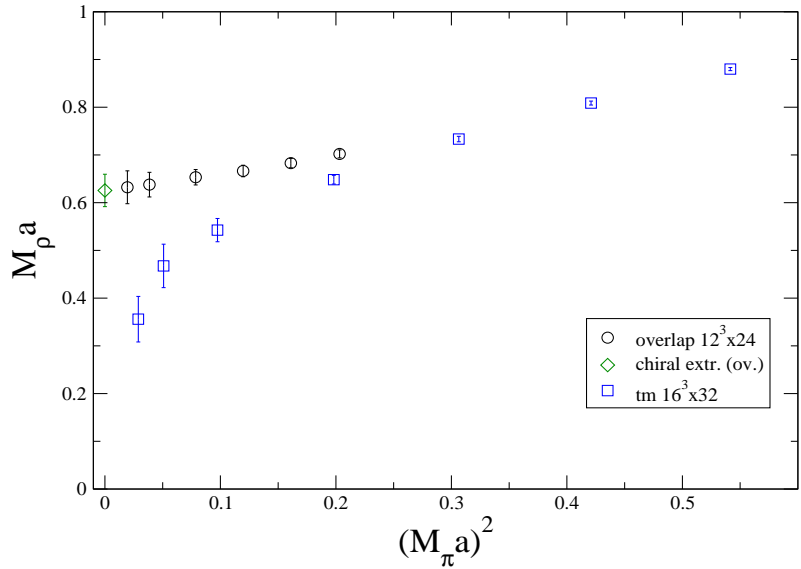}
\end{center}
\vspace{-0.0cm}
\caption{\it Comparison of results for the vector meson mass as a function
of the pion mass squared for overlap and twisted mass fermions. The
chiral extrapolation in the overlap case is done with a linear fit in 
$M_\pi^2$.
\label{fig:mrho_ov_tm}}
\end{figure}

\begin{figure}
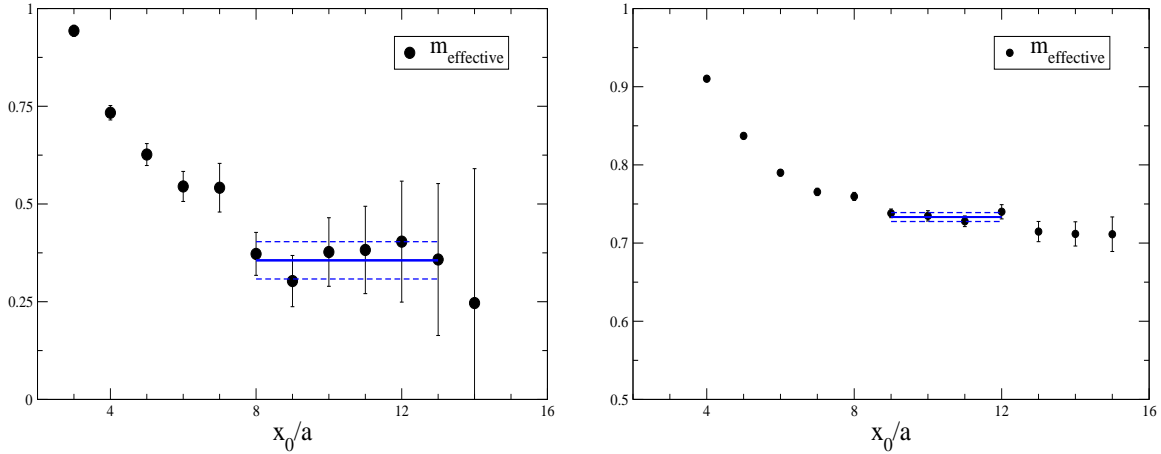

\vspace{+0.7cm}
\hspace*{-0.4cm}
\epsfig{file=figure/axve1_0_emass.eps,width=7.3cm,height=6.0cm}
\put(20,0){\epsfig{file=figure/axve1_4_emass.eps,width=7.3cm,height=6.0cm}}
\vspace{-0.0cm}
\caption{\it Effective mass plateaux for the vector meson mass at $\mu
a=0.005$ and at $\mu a=0.06$. Notice the different scale on the $y$ axis.
\label{fig:axve1_0_emass}}
\end{figure}

In order to extract a reliable value for the vector meson mass, 
sink-smearing has been used. For overlap data, where the volume is not
so large and the statistical errors are consequently significant, the
smearing procedure is important in order to isolate the ground state 
(at the same time reducing the statistical error of the
estimated mass). For twisted mass data on the largest volume
($16^3\times32$), smearing has a small effect, apparently because
it is not able to decrease substantially the coupling of the 
interpolating operators to the excited states. 
The vector meson mass has been extracted by averaging the values of the
effective mass within a time interval which excludes the points
at the largest times. This points are in fact affected by large 
statistical uncertainty
and tend to lower the estimated value. The results are reported 
in Table~\ref{tab:mmesons3} and plotted in Figure~\ref{fig:mrho_ov_tm}.

In the case of Wtm fermions, we report in 
Fig.~\ref{fig:axve1_0_emass} two examples for the plateaux of the 
effective mass of the vector meson, namely the case corresponding to the
smallest quark mass $\mu a=0.005$ (which has highest statistical
fluctuations and for which we have chosen the plateau
in the range $8\le x_0\le13$) and that of an intermediate mass 
$\mu a=0.06$ (with plateau chosen in the range $9\le x_0\le12$). 
Despite the precautions described above, we still observe a
strong decrease of the vector meson mass at low quark masses, as compared to
the overlap case where the behavior with the pion mass squared is linear
to a good approximation. We will discuss later, when examining
$f_\pi$, the problems that can arise in the Wtm case when 
going down to low masses at fixed
lattice spacing.  
This phenomenon has to be further investigated, for example by 
going to higher $\beta$ values.
%or by studying correlation matrices of a larger set of operators. 
%{\bf Can this be done already for this paper?}
This is beyond the scope of the present work
and will be addressed in another (presently on-going) project whose 
ultimate goal is the study of
the scaling behavior of various quantities in twisted mass QCD computed
for a wide range of quark masses~\cite{on_going_scaling}. 
    
\subsection{Renormalization constants}
Using the method explained in
Refs.\ \cite{HJLW01,HJLWproc} we computed the 
renormalization group invariant (RGI) quark mass renormalization constant
$Z_m^{\textrm{RGI}}$ for Wtm and overlap fermions. 
Here we just describe the method and we refer to Refs.\ \cite{HJLW01,HJLWproc}
for a more detailed explanation.
There are essentially two requirements for appling this method: \\
\vspace{-0.5cm}
\begin{itemize}
\item Due to the symmetries of the lattice action, 
renormalization constants of different local operators
are related in a simple way. \\
\vspace{-0.5cm}
\item Using an alternative discretization of QCD it is possible to compute 
a renormalized 
matrix element (or quark mass) in the continuum. \\
\end{itemize}
\vspace{-0.5cm}
The basic idea is to compute a universal factor --- that could be a RGI matrix 
element or quark mass --- in the continuum at a fixed reference value 
of a physical quantity (for example the pion mass). 
Then matching this universal factor with a renormalized matrix element 
(the bare matrix element computed in the target regu\-larization) at the 
reference point
and at fixed lattice spacing, it is trivial to extract the renormalization factor.
Obviously the method works, using the symmetry properties of the target 
regularization, 
if it is possible to relate renormalization factors of different local operators.
To this end, we have used two different matching conditions. We have matched 
the RGI quark mass (method 1) and the matrix element of the pseudoscalar 
density (method 2) (see \cite{HJLWproc} for details) at the reference
points given by $x_{\rm ref} = (r_0M_\pi)^2=1.5736,~3.0,~5.0$, where the
last point is considered only for twisted mass fermions (in the overlap 
case we do not have data in the region corresponding to such high pion
masses).
The universal factor was obtained using the renormalization constants 
and the bare matrix elements computed by the ALPHA collaboration 
\cite{ALPHA_mass,Garden} using ${\cal O}(a)$ non-perturbatively improved
Wilson fermions.
For overlap fermions, due to the lattice chiral symmetry \cite{ML},
we have the following relation between the renormalization factors of 
the pseudoscalar and scalar density and the renormalization 
factor of the quark mass,
\be
Z_P = Z_S = \frac{1}{Z_m} \ .
\ee
For Wtm, due to the existence of an exact flavor symmetry for 
massless Wilson quarks, we obtain 
\be
Z_P = \frac{1}{Z_{\mu}}
\ee 
for all the flavor components.
We summarize our results in Table \ref{tab:Z}.  
\begin{table}[t]
\begin{center}
\begin{tabular}{l l l l l}
\hline
\hline
& method 1  &  method 2  & method 1 & method 2 \\
$x_{\rm ref}$ & $Z_m^{\rm RGI,ov}$  &  $Z_m^{\rm RGI,ov}$ &  
$Z_{\mu}^{\rm RGI,tm}$ & $Z_{\mu}^{\rm RGI,tm}$      \\
\hline
1.5736  & 1.02(6)  & 0.98(5)  &  2.27(7) &  2.22(8)      \\
3.0     & 0.98(7)  & 1.01(5)  &  2.32(6) &  2.36(6)      \\
5.0     & --       & --       &  2.39(5) &  2.55(20)     \\
\hline
\hline
\end{tabular}
\caption{Results for $Z_m^{\rm RGI,ov}$ and $Z_{\mu}^{\rm RGI,tm}$. \label{tab:Z}}
\end{center}
\end{table}
For overlap fermions, these results are basically independent from 
choosing $x_{\rm ref}=1.5736,3.0$. In the twisted mass case, the slight
dependence (always within 1-2 standard deviations) is due to the 
${\cal O}(a^2\mu^2)$ lattice artifacts that affect the quantities used for the
matching. The large error on $Z_{\mu}^{\rm RGI,tm}$ at 
$x_{\rm ref}=5.0$ from method 2 comes from the error on the 
universal factor 
computed from the data of the ALPHA collaboration in the continuum limit.
The rather large value of the renormalization constant 
in the case of Wtm fermions 
is reflected in the slope of the curves shown 
in Figures \ref{fig:mpi_ov_tm} and \ref{fig:mpi_tm_lin_quad}. 

\subsection{Ward identities quark masses}
\label{sec:pcac}

\begin{figure}
\vspace{-0.0cm}
\begin{center}
\epsfig{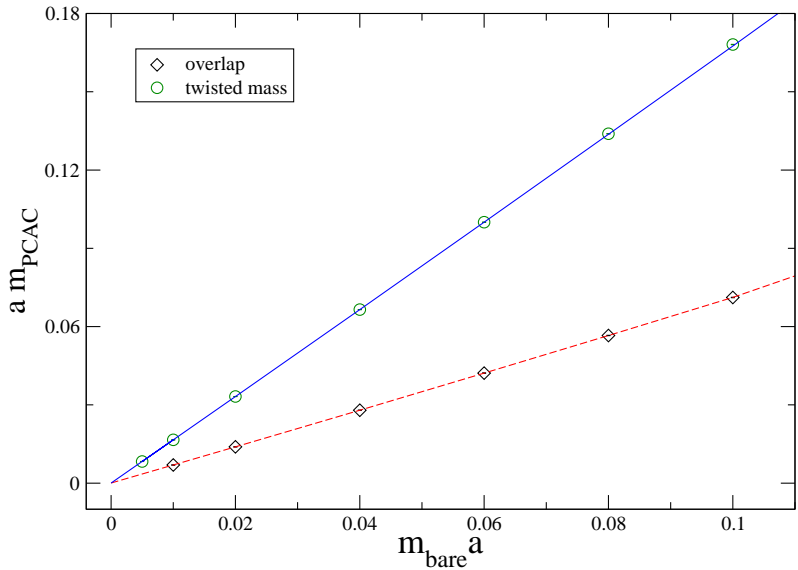}
\end{center}
\vspace{-0.0cm}
\caption{\it Comparison of results for the WI quark masses as a 
function of the quark mass for overlap and twisted mass 
fermions. The values of $m_{\textrm{PCAC}}$ and $m_{\textrm{PCVC}}$
extrapolated quadratically to 
the chiral limit are $0.00004(2)$ and $0.00003(1)$ for the overlap and the
Wtm case respectively.
\label{fig:mpcac}}
\end{figure}

The Ward identities (WI) quark masses $m_{\textrm{PCAC}}^{\textrm{ov}}$ and 
$m_{\textrm{PCVC}}^{\textrm{tm}}$ can be extracted from the ratios
\begin{eqnarray} \nonumber
m_{\textrm{PCAC}}^{\textrm{ov}} & = & \frac{\sum_{\vec x}\langle \partial_0 
A_0^\dag(\vec x , x_0 )\; 
P(0)\rangle}{2\sum_{\vec x}\langle P^\dag(\vec x , x_0 ) P(0)\rangle} \ , \\
m_{\textrm{PCVC}}^{\textrm{tm}} & = & \frac{\epsilon^{3bc}\sum_{\vec x}\langle
\partial_0 V_0^b(\vec x , x_0)\; P^c(0)\rangle}{2\sum_{\vec x}\langle 
P^c(\vec x , x_0) P^c(0)\rangle} \ .
\end{eqnarray}
Results for the (WI) quark masses are reported in Table \ref{tab:mpcac} 
and plotted in Figure \ref{fig:mpcac}, together with the results
of a quadratic extrapolation to the chiral limit.

\begin{table}
\begin{center}
\begin{tabular}{ccc}
\hline
\hline
$m_{\rm bare} a$& $m_{\textrm{PCAC}}^{\textrm{ov}} a $ &
$m_{\textrm{PCVC}}^{\textrm{tm}} a$ \\
\hline
0.005 &     -    &0.008303(7)\\
0.01 & 0.00695(2)&0.016602(8)\\
0.02 & 0.01391(3)&0.033187(17)\\
0.04 & 0.02795(5)&0.066514(24)\\
0.06 & 0.04218(6)&0.100051(35)\\
0.08 & 0.05659(7)&0.133899(42)\\
0.10 & 0.07116(8)&0.168110(51)\\
\hline
\hline
\end{tabular}
\end{center}
\caption{\it Ward identities quark masses. \label{tab:mpcac}}
\end{table}

\begin{figure}
\vspace{-0.0cm}
\begin{center}
\epsfig{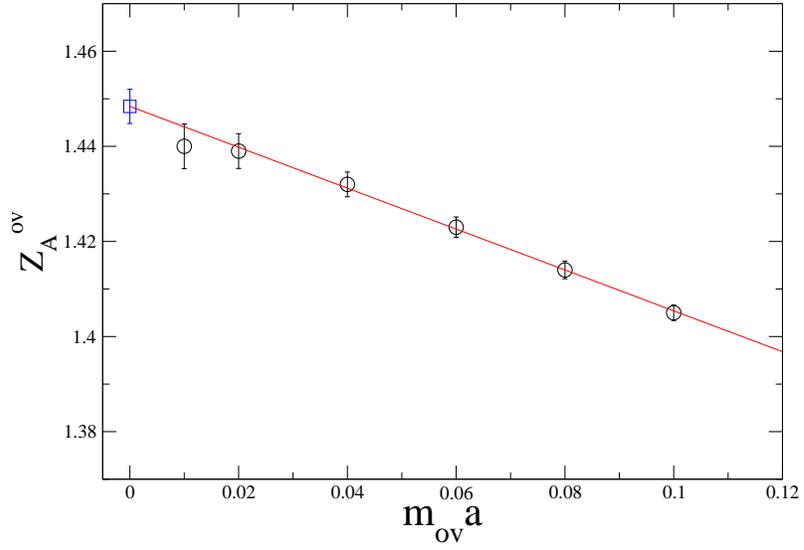}
\end{center}
\vspace{-0.0cm}
\caption{\it $Z^{ \rm ov}_A$ as function of the quark mass and its chiral
  extrapolation.\label{fig:zetaA} }
\end{figure}

\begin{figure}
\vspace{-0.0cm}
\begin{center}
\epsfig{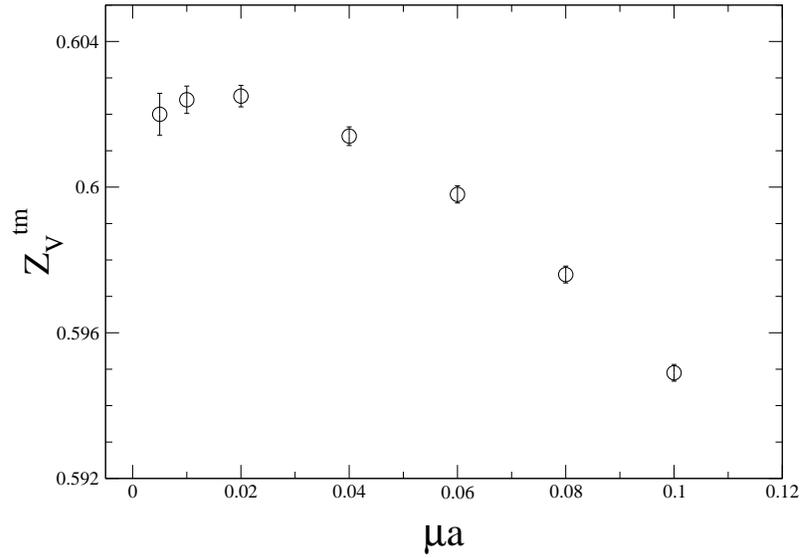}
\end{center}
\vspace{-0.0cm}
\caption{\it $Z^{\rm tm}_V$ as function of the quark mass.\label{fig:zetaA_tm}}
\end{figure}

From the values of the WI quark masses and the bare quark masses, the 
corresponding renormalization factors can be computed at each value of 
the bare quark mass. They are then given by
\be
Z^{\textrm{ov}}_A=\frac{m_{\rm ov}}{m_{\textrm{PCAC}}^{\textrm{ov}}} \ , \qquad
Z^{\textrm{tm}}_V=\frac{\mu}{m_{\textrm{PCVC}}^{\textrm{tm}}}\; .
\label{eq:zetas}
\ee
The behavior of the WI quark masses in Figure \ref{fig:mpcac} 
is only apparently linear. In the overlap case this behavior is well described
by a quadratic curve (where the quadratic term is quite small). 
This is reflected in the linear behavior of $Z^{\textrm{ov}}_A$
with a rather mild slope (see Figure \ref{fig:zetaA}). A linear fit 
(excluding the data point at $am_{\rm ov} = 0.01$) allows then 
to obtain a value for $Z^{\textrm{ov}}_A$ in the 
chiral limit, $Z^{\textrm{ov}}_A=1.448(4)$. In the twisted mass
case even a quadratic fit is inadequate to describe the behavior of
the PCVC mass with respect to the bare mass $\mu$. This is reflected in
the behavior of $Z^{\textrm{tm}}_V$ as function of $\mu$ --- see 
Figure \ref{fig:zetaA_tm} --- which, in particular at small quark masses, 
is altered by residual lattice artifacts. This affects
the denominator of Eq.~(\ref{eq:zetas}) 
(i.e. $m_{\textrm{PCVC}}^{\textrm{tm}}$), giving rise to a strong 
non-linearity. For this reason we prefer not to quote any value for 
$Z^{\textrm{tm}}_V$ in the chiral limit.
%we do not include the lowest
%three points in the chiral extrapolation. In the chiral limit we get
%$Z^{\textrm{tm}}_V=0.6061(3)$ (only the statistical error is quoted).

\subsection{Pseudoscalar decay constants}
\label{sec:decay}

The pseudoscalar decay constants can be computed from the ratios 
\be
f^{\textrm{ov}}_\pi=\frac{Z^{\textrm{ov}}_A |\langle 0|A_0|\pi\rangle_{\rm ov}|}{M_\pi^{\rm ov}}\;,\qquad
f_\pi^{\textrm{tm}}=\frac{Z^{\textrm{tm}}_V |\langle 0|V_0^b|\pi\rangle_{\rm tm}|}{M_\pi^{\rm tm}}\qquad b=1,2
\ee
which require the determination of $Z^{\textrm{ov}}_A$ and of 
$Z^{\textrm{tm}}_V$ as discussed in the previous section (where 
at maximal twist $Z^{\textrm{tm}}_V$ should be identical to the 
value computed with standard Wilson fermions).

There is also a
second, ``indirect'' method that uses the PCAC and PCVC relations and does not 
require the computation of any renormalization constant,
\be
f^{\textrm{ov}}_\pi = \frac{2m_{\rm ov}}{(M_\pi^{\rm ov})^2} |
\langle 0|P|\pi\rangle_{\rm ov}| \ , \quad
f_\pi^{\textrm{tm}} = \frac{2\mu}{(M_\pi^{\rm tm})^2} | \langle
0|P^b|\pi\rangle_{\rm tm}| \ , \quad b=1,2 \ .
\label{indirect}
\ee
This second method will prove to be useful especially for the Wtm case. 

\begin{figure}
\vspace{-0.0cm}
\begin{center}
\epsfig{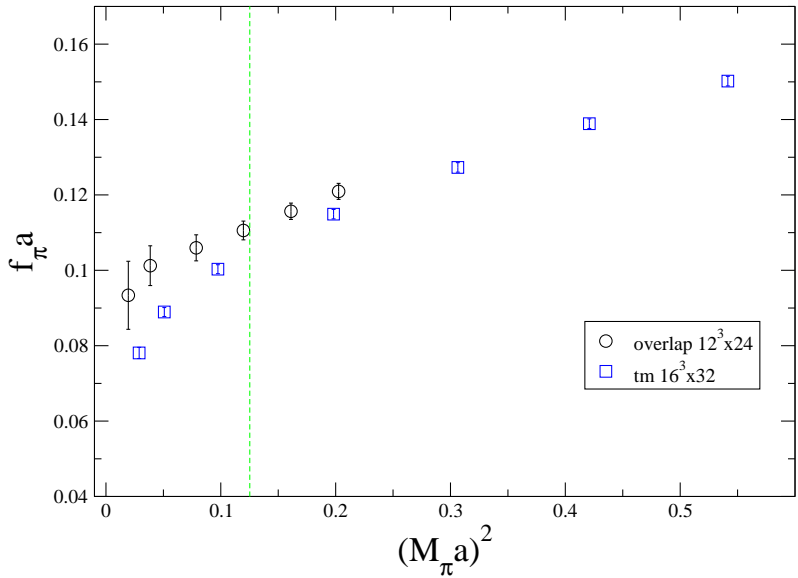}
\end{center}
\vspace{-0.0cm}
\caption{\it Comparison of results for the pion decay constant as a 
function of the pion mass squared for overlap and twisted mass 
fermions. The vertical line represents the value of $(M_\pi a)^2$ which 
corresponds to a bare quark mass $\bar m_q a=a^2 \Lambda_{\rm
QCD}^2$, i.e. the r.h.s. of Eq.~(\ref{COND1}) apart from an 
unknown proportionality factor.
\label{fig:fpi_ov_tm_final}}
\end{figure}

From the theoretical side we expect from 
one  loop quenched chiral perturbation theory (q$\chi$PT) that 
$f_\pi$ has neither chiral logarithms nor finite volume effects,
\be  \label{fpif}
f_\pi=f\left(1+\frac{\alpha_5}{(4 \pi f)^2} M_\pi^2\right) \ ,
\ee
where $\alpha_{5}$ is a low energy constant.

\begin{table}
\begin{center}
\begin{tabular}{ccc}
\hline
\hline
$m_{\rm bare} a$& $f^{\textrm{ov}}_{\pi} a$ &
$f^{\textrm{tm}}_{\pi} a$ \\
\hline
0.005 &  -       &0.0781(15)\\
0.01 & 0.0934(90)&0.0889(12)\\
0.02 & 0.1012(53)&0.1003(12)\\
0.04 & 0.1060(34)&0.1149(12)\\
0.06 & 0.1106(25)&0.1273(13)\\
0.08 & 0.1157(22)&0.1389(13)\\
0.10 & 0.1209(21)&0.1502(13)\\
\hline
\hline
\end{tabular}
\end{center}
\caption{\it Pseudoscalar decay constants from the ``indirect'' method,
Eqs.~(\ref{indirect}).\label{tab:decay_const}}
\end{table}

As we have seen in the previous section, a reliable number for 
$Z^{\textrm{tm}}_V$ cannot be provided given the strong quark mass dependence
of the renormalization constant in the Wtm case. 
We therefore only give values for $f_\pi$ from the ``indirect'' method.
Results for the pion decay constant obtained with the ``indirect'' method 
for both overlap (from $C_{P-S,\rm ov} (x_0)$) and
twisted mass fermions (from $C_{P,\rm tm}^b (x_0)$) 
are reported in Table \ref{tab:decay_const} and 
plotted in Figure \ref{fig:fpi_ov_tm_final}. 

This Figure reveals that in the case of the overlap fermion $f_\pi$ nicely 
follows the linear behavior predicted from quenched chiral perturbation
theory. However, in the case of Wtm fermions we observe a {\em bending} of 
$f_\pi^\mathrm{tm}$ when the pion mass is small. 
It has been argued \cite{Frezzotti:2003ni} that the condition
\begin{equation}\label{COND1}
m_q a\, \gg \, a^2 \Lambda_{\rm QCD}^2 \, 
\end{equation}
(where $m_q=\sqrt{\tilde m^2+\mu^2}$ is equal to $\mu$ at full twist) 
has to be satisfied (with some proportionality 
factor in front of the r.h.s.) 
in order the explicit breaking of the chiral
symmetry to be driven by the the mass 
term rather than by the Wilson term (in which case large 
cut-off effects could appear). If this is the effect we are 
seeing in Figure \ref{fig:fpi_ov_tm_final}, then this inequality seems 
to be satisfied with a 
proportionality constant of $\mathcal{O}(1)$. 
In order to visualize this, we represent the squared pion mass
corresponding to a bare quark mass $\bar m_q a=a^2 \Lambda_{\rm QCD}^2$
%(i.e. the r.h.s. of Eq.~(\ref{COND1}) without the proportionality 
%coeefficient) 
as the vertical dotted line in Figure \ref{fig:fpi_ov_tm_final}.
What is puzzling is that, for a setup where 
the correlators are $\mathcal{O}(a)$ improved (as it should be in our 
simulations of 
Wilson twisted mass fermions), one could hope that the 
condition that has to be satisfied is
$m_q a\, \gg \, a^3 \Lambda_{\rm QCD}^{3}$ rather than the condition 
in Eq.~(\ref{COND1}).  
In this case, all of our data should be safe, unless the proportionality
constants in front of the r.h.s. is a large number.  
Clearly the bending phenomenon observed in 
Figure \ref{fig:fpi_ov_tm_final} deserves further analytical and
numerical investigations and can be clarified presumably only when 
results at smaller lattice spacings are available. 

For overlap fermions, the renormalization constant can be reliably extracted
in the chiral limit and hence also the ``direct'' method can be used.
For these two methods we have considered both the simplest correlators
$C_{P,\rm ov}(x_0)$ (``indirect'') and $C_{A_0,\rm ov}(x_0)$ (``direct''),
as well as the composite ones 
$C_{P-S,\rm ov}(x_0)$ (``indirect'') and $C_{A_0+V_0,\rm ov}(x_0)$ (``direct''),
where the finite volume effects from the zero modes cancel. The results in the
chiral limit for the
four cases turn out to be compatible within the errors,
\bea
\lim_{m_{\rm ov}\rightarrow 0} f^{\textrm{ov}}_{C_{P-S}} a &=& 0.0963(52)\ , \nn\\ 
\lim_{m_{\rm ov}\rightarrow 0} f^{\textrm{ov}}_{C_P} a &=& 0.0980(46)\ , \nn\\
\lim_{m_{\rm ov}\rightarrow 0} f^{\textrm{ov}}_{C_{A_0+V_0}} a &=& 0.1010(40)\ , \nn\\
\lim_{m_{\rm ov}\rightarrow 0} f^{\textrm{ov}}_{C_{A_0}} a &=& 0.1018(38) \;.\nn
\eea
This means that the finite volume effects, which in the
case of $C_{P,\rm ov}(x_0)$ and $C_{A_0,\rm ov}(x_0)$ may affect both the
pseudoscalar mass and the matrix element, cancel out when one takes
the suitable ratio needed to compute $f_\pi$. The slight difference between 
the ``direct'' and the ``indirect'' method can be ascribed to the 
different $\mathcal{O}(a^2)$ lattice
artifacts of the two correlators used and also to the uncertainty in
the chiral extra\-polation of $Z^{\textrm{ov}}_A$ needed for 
the ``direct'' method.  

\begin{figure}[t]
\vspace{-0.0cm}
\begin{center}
\epsfig{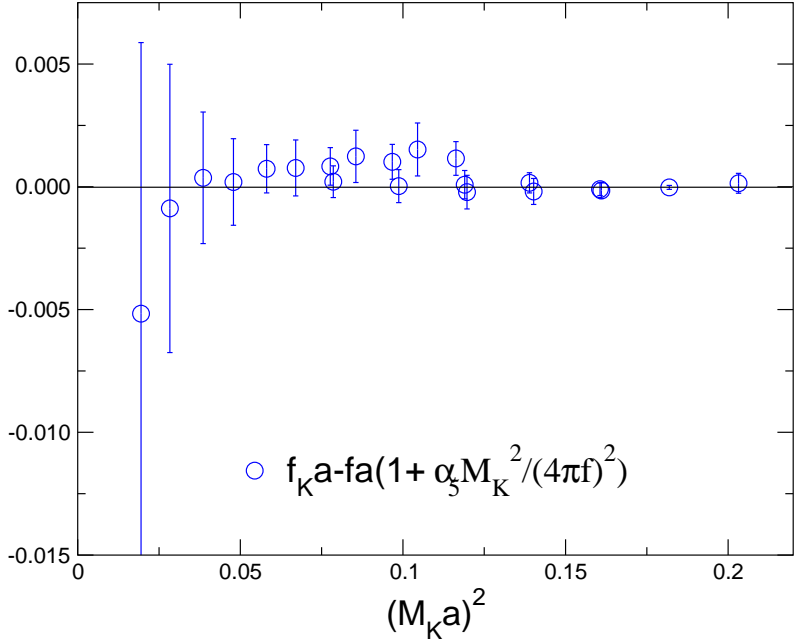}
\end{center}
\vspace{-0.0cm}
\caption{\it Difference between the results for the decay constant and a
linear fit to degenerate data only. \label{fig:fk_ndeg_PP-SS}}
\end{figure}

Since for overlap fermions the behavior of $f^{\textrm{ov}}_\pi$ is perfectly linear, as
predicted by quenched chiral perturbation theory --- 
in contrast to the Wtm case --- we can hope to check 
the prediction of q$\chi$PT in the
case of $f^{\textrm{ov}}_K$. Moreover, since the two methods for
extracting $f_\pi$ are in good agreement, in the non-degenerate case 
we use only the PCAC method,
\be
f^{\textrm{ov}}_K=\frac{m_{{\rm ov},1}+m_{{\rm ov},2 }}{(M_K^{\rm ov})^2}|\langle 0|P|K\rangle_{\rm ov}|\;.
\ee 
At one loop in q$\chi$PT, $f_K$ takes the form
\be
f_K=f\left(1+\frac{\alpha_5}{(4 \pi f)^2} M_K^2+ \textrm{FV}
+ \textrm{LG}\right)
\ee
where $f$ and $\alpha_5$ are the same as in Eq.~(\ref{fpif}), ``FV'' are finite size effects
and ``LG'' logarithmic corrections (see Ref. \cite{becirevic_villadoro} for the complete formula). 

One could envisage the following strategy: determine $f$ and 
$\alpha_5$ from the degenerate data and search for FV and LG.
Unfortunately, within our statistical accuracy 
we see only a linear behavior in $M_K^2$, with $\alpha_5$ in perfect 
agreement with the
determination from the degenerate data (i.e.\ from $f_\pi$). 
This is displayed in Figure~\ref{fig:fk_ndeg_PP-SS}, where we plot 
the difference between the data and the values obtained from the
q$\chi$PT formula for $f_K$ without the terms FV and LG and where
$\alpha_5$ and $f$ are determined from the degenerate case only.
 
The results we get are $f_\pi=155(11)$ MeV, $f_K=173(8)$ MeV, 
$f_K/f_\pi=1.11(3)$, $\alpha_5 = 1.85(30)$. As usual in the 
quenched approximation, the
value of $f_K/f_\pi$ turns out to be about $10\%$ smaller than its
experimental value.

\subsection{Baryon masses}

In order to extract baryon masses we use the following interpolating
operators (for the octet and the decuplet respectively),
\begin{eqnarray}
B_\alpha^{\textrm{oct}}&=&\epsilon^{ABC}\left[((d^A)^T C\gamma_5
u^B)u^C_\alpha-((u^A)^T C\gamma_5 d^B)u^C_\alpha\right] \ , \nonumber \\
B_{k,\alpha}^{\textrm{dec}}&=&\epsilon^{ABC}((u^A)^T C\gamma_k
u^B)u^C_\alpha \ , \quad k=1,2,3 \ ,
\end{eqnarray}
where $C$ is the charge conjugation matrix.
For the decuplet $k=1,2,3$ are equivalent. We have chosen 
$k=1$.

For correlators at zero momentum in the overlap case, we have
\be
\sum_{\vec x}(1+\gamma_4)_{\alpha\beta} \langle \bar
B_\alpha^{\textrm{oct,dec}}(\vec x , x_0 )
B_\beta^{\textrm{oct,dec}}(0)\rangle_{\rm ov}
\propto e^{-M x_0} \ , \quad a \ll x_0 \leq \frac{T}{2}\;.
\ee 
In this case we perform a simple exponential fit in the first half 
of the lattice in the time direction to avoid contaminations
coming from the state with opposite parity. 
In the twisted mass case (at twist angle $\omega=\pi/2$) it is easy to show that
\be
\langle \bar
B_\alpha^{\textrm{oct,dec}}(x)B_\beta^{\textrm{oct,dec}}(0)\rangle_{\textrm{phys}}=
\frac{1}{2}(1+i\gamma_5)_{\alpha\gamma}\langle
\bar B_\gamma^{\textrm{oct,dec}}(x)B_\delta^{\textrm{oct,dec}}(0)\rangle_{\textrm{tm}}(1+i\gamma_5)_{\delta\beta}\;,
\ee 
where $\langle \bar B_\alpha^{\textrm{oct,dec}}(x)
B_\beta^{\textrm{oct,dec}}(0)\rangle_{\textrm{phys}}$
is the correlator with the correct quantum numbers in the continuum.
 
\begin{figure}[t]
\vspace{-0.0cm}
\begin{center}
\epsfig{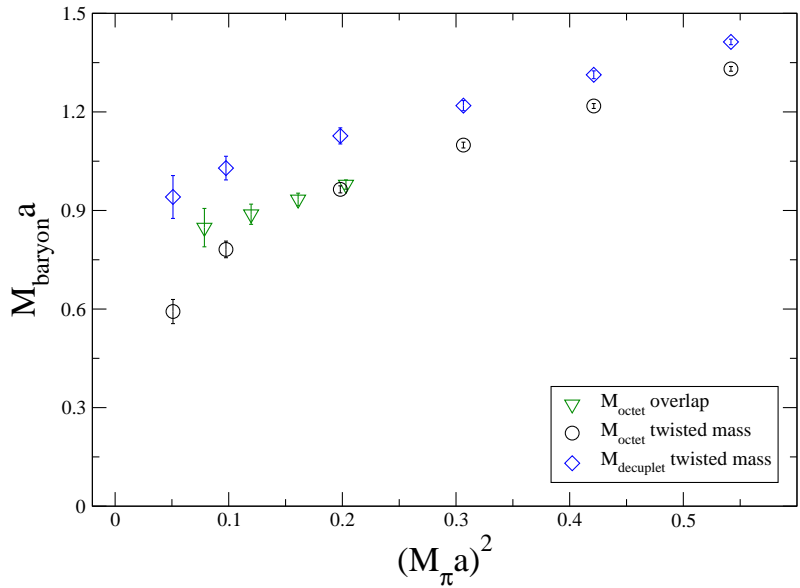}
\end{center}
\vspace{-0.0cm}
\caption{\it Comparison of results for the baryon masses (octet and decuplet
  in the SU(3) symmetric limit) as a 
function of the pion mass squared for overlap and twisted mass 
fermions. \label{fig:mhadrons}}
\end{figure}

Results (obtained by using sink-smearing) are reported in 
Table \ref{tab:mhadrons} and plotted in Figure \ref{fig:mhadrons}. 
Concerning overlap fermions, due to the relatively small
volume, the decuplet channel as well as the octet correlators
corresponding to the lowest two bare quark masses are
too noisy and we are not able to extract the corresponding masses.
In the Wtm case, we observe a bending of the data at small quark
masses similar to the case of the vector meson, whereas at bare quark masses
larger than 50 MeV the results seems to behave very similarly to the 
overlap data.

\begin{table}[h]
\begin{center}
\begin{tabular}{cccc}
\hline
\hline
$m_{\rm bare} a$& $M^{\textrm{ov}}_{\textrm{octet}} a$
& $M^{\textrm{tm}}_{\textrm{octet}} a $ & 
$M^{\textrm{tm}}_{\textrm{decuplet}} a$\\
\hline
0.01 &   -       & 0.592(37) & 0.941(65)\\
0.02 &   -       & 0.782(25) & 1.029(36) \\
0.04 & 0.8479(58)& 0.963(11) & 1.126(25) \\
0.06 & 0.8885(31)& 1.0987(88)& 1.2199(15) \\
0.08 & 0.9336(19)& 1.2183(74)& 1.3124(12)  \\
0.10 & 0.9788(14)& 1.3302(72)& 1.4136(83) \\
\hline
\hline
\end{tabular}
\end{center}
\caption{Baryon masses with sink smearing.\label{tab:mhadrons}}
\end{table}

\section{Conclusions}

In this paper we confronted quenched simulation results of overlap fermions 
and Wtm fermions in their approach to the chiral limit. 
We emphasize that we tested both lattice discretizations of QCD at only
one value of $\beta=5.85$ corresponding to a value of the lattice spacing
of $a=0.123$ fm. Although scaling tests are certainly of great importance
to further explore the potential of both kind of fermions, our results 
show already very interesting features. 

The first is that indeed with both kind of lattice fermions small values
of the light meson masses can be reached, such as $M_\pi \simeq 220$ MeV
in the overlap case and $M_\pi\simeq 270$ MeV in the Wtm case.
In addition, the statistical fluctuations for the observables studied 
here are comparable for both formulations.
This is very promising. In a detailed algorithmic study 
\cite{solver} we find that Wilson twisted mass fermions are a factor
of 20 to 40 faster than overlap fermions. 
Thus Wilson twisted mass fermions
have the potential for dynamical fermions simulations at realistically
small quark masses on the next generation of supercomputers in the 
multi-teraflops range. 

%Nevertheless, the presence of such a first order phase transition 
%renders the approach to the chiral limit in full QCD difficult.

However, we believe that a number of questions have to be addressed
to understand better the Wtm formulation of lattice QCD:
In the quenched comparison of overlap and Wilson twisted mass fermions,
we encountered a {\em ``bending'' phenomenon} for Wtm fermions. This effect 
manifests itself in the chiral approach of all the quantities
studied here. While the data for overlap fermions
extrapolate nicely linearly, the data for Wtm fermions show a bending when the
quark mass assumes too small values. 
This effect might be explained by the interplay of the Wilson term and the
twisted mass term, which requires one of the two inequalities discussed in
Sec.~\ref{sec:decay} to be satisfied.

For the present simulations at only one value of the lattice spacing we 
are not able to determine the cause of this ``bending'' phenomenon 
and whether it 
disappears in the continuum limit. For this, a detailed scaling analysis 
would be necessary, a work that is in progress. 
Overlap fermions on the other hand nicely approach the chiral limit 
close to the physical point with realistic light meson masses. 
It seems that the conceptual virtues of this approach become more 
and more important, the smaller the quark mass is chosen. 
However, we believe that a final answer, which formulation to use 
for extracting physics in the chiral limit, can only be given when 
the scaling behavior of both approaches is understood. 

For dynamical simulations the presence of a first order phase transition
has been seen for Wtm fermions \cite{tmdyn}.              
This observation is in accordance with an
effective potential picture 
\cite{SHARPE-SINGLETON,MUNSTER,Sharpe:2004ps,Luigi}.
The findings in Ref.~\cite{tmdyn} just emphasize the fact that the study
of the phase structure of lattice QCD is a necessary prerequisite for reliable
physics results. 

\subsection*{Acknowledgments}
We are indebted to 
R. Frezzotti and G. C. Rossi for many useful remarks and suggestions.
We thank also M. M\"uller-Preu\ss ker for useful discussions.
The computer centers at NIC/DESY Zeuthen, NIC at the Forschungszentrum 
J\"ulich and HLRN provided the necessary technical 
help and the computer resources. 
We thank the staff of these centers for their help and advise.
This work was supported
by the DFG Sonderforschungsbereich/Transregio SFB/TR9-03.

%\newpage

\end{document}